\documentclass[aps, reprint, amsfonts, amssymb, superscriptaddress]{revtex4-1}
\usepackage{hyperref, graphicx, wrapfig, amsmath, color}
\newcommand{\be}{\begin{equation}}
\newcommand{\ee}{\end{equation}}
\newcommand{\bea}{\begin{eqnarray}}
\newcommand{\eea}{\end{eqnarray}}
\begin{document}
\title{Quadratic constrained mixed discrete optimization with an adiabatic quantum optimizer}

\author{Rishabh Chandra}
\affiliation{Department of Mechanical Engineering, Purdue University, West Lafayette, Indiana 47907, USA}
\author{N. Tobias Jacobson}
\email{ntjacob@sandia.gov}
\affiliation{Sandia National Laboratories, Albuquerque, New Mexico 87185, USA}
\author{Jonathan E. Moussa}
\affiliation{Sandia National Laboratories, Albuquerque, New Mexico 87185, USA}
\author{Steven H. Frankel}
\affiliation{Department of Mechanical Engineering, Purdue University, West Lafayette, Indiana 47907, USA}
\author{Sabre Kais}
\affiliation{Departments of Chemistry and Physics, Purdue University, West Lafayette, Indiana 47907, USA}
\affiliation{Qatar Environment and Energy Research Institute, Qatar Foundation, Doha, Qatar}

\begin{abstract}
We extend the family of problems that may be implemented on an adiabatic quantum optimizer (AQO). 
When a quadratic optimization problem has at least one set of discrete controls and the constraints are linear, we call this a quadratic constrained mixed discrete optimization (QCMDO) problem. 
QCMDO problems are NP-hard, and no efficient classical algorithm for their solution is known. 
Included in the class of QCMDO problems are combinatorial optimization problems constrained by a linear partial differential equation (PDE) or system of linear PDEs. 
An essential complication commonly encountered in solving this type of problem is that the linear constraint may introduce many intermediate continuous variables into the optimization while the computational cost grows exponentially with problem size. 
We resolve this difficulty by developing a constructive mapping from QCMDO to quadratic unconstrained binary optimization (QUBO) such that the size of the QUBO problem depends only on the number of discrete control variables. 
With a suitable embedding, taking into account the physical constraints of the realizable coupling graph, the resulting QUBO problem can be implemented on an existing AQO. 
The mapping itself is efficient, scaling cubically with the number of continuous variables in the general case and linearly in the PDE case if an efficient preconditioner is available.
\end{abstract}

\maketitle

\section{Introduction}
Quadratic unconstrained binary optimization (QUBO) is the set of problems for which the objective functional is quadratic in binary variables that are otherwise unconstrained. 
QUBO is NP-hard from a computational complexity perspective \cite{Barahona1982}. 
A number of interesting problems can be mapped to QUBO form, including those from the fields of image recognition \cite{Neven2008}, machine learning \cite{Neven2008B}, protein folding \cite{Perdomo-Ortiz2012}, and number theory \cite{Gaitan2012, Bian2013}. 
As the membership of useful problems in QUBO has grown, interest has intensified to develop QUBO solvers that can practically solve problems of increasing complexity. 
Adiabatic quantum optimization (AQO) is an alternative to standard classical heuristic algorithms for solving QUBO problems. 
In AQO, or quantum annealing, a system is initialized into the easily-prepared ground state of some initial Hamiltonian $H_{0}$, and then this Hamiltonian is slowly distorted into a final problem Hamiltonian $H_{1}$ \cite{Finnila1994, Kadowaki1998}. 
The Hamiltonian $H_{1}$ is constructed such that its ground state corresponds to the solution of an optimization problem of interest. 
By the adiabatic theorem, if the Hamiltonian is modified sufficiently slowly, the system will remain at all times in the ground state of the instantaneous Hamiltonian  \cite{Lidar2009}. 
At the conclusion of the interpolation, the system can be measured to read out the solution. 
For certain problems, quantum annealing is known to provide a speedup over classical algorithms \cite{Somma2012}, but the extent of problems for which quantum annealing performs faster than any known classical algorithm is still unknown. 
Recently, considerable attention has been devoted to answering whether an AQO can provide a speedup for solving QUBO \cite{Boixo2013b, Boixo2014, McGeoch2013}. 
An AQO platform is currently available on which QUBO problems can be implemented \cite{Johnson2011}. 
Though it is currently an open question whether such an AQO platform will provide a qualitative and scalable speedup in solving QUBO over classical algorithms, developing new problems that such a device can implement motivates research in this area.

In this work, we consider the set of problems where the objective function is quadratic, the constraints are linear, some of the controls are discrete, and other controls may be continuous. 
We call this quadratic constrained mixed discrete optimization (QCMDO). 
QCMDO problems appear in a number of contexts, notably where the constraints are given by a linear partial differential equation (PDE) or system of linear PDEs, including gas/water network flow optimization \cite{Martin2006, Hante2009, Geissler2011}, traffic optimization \cite{Fugenschuh2006}, and microchip cooling optimization \cite{Xu2007}. 
QCMDO is also NP-hard, as it contains QUBO. 
Linearly-constrained problems with only continuous controls are tractable because of their convex structure. 
However, the discrete nature of the controls in QCMDO destroys convexity. 

We show that QCMDO can be mapped efficiently into quadratic unconstrained discrete optimization, which in some cases may then be efficiently mapped to QUBO. 
This mapping adds to the family of interesting problems that may be implemented on an AQO for which the problem Hamiltonian takes the form of a classical Ising model. 
If indeed an AQO were to provide a speedup over classical QUBO solvers, this speedup would translate directly to faster solution of QCMDO problems as well. 
However, notwithstanding such a speedup this mapping may serve as a useful method of casting the problem for standard classical solvers as well, since the dimensional reduction of the mapping efficiently removes a potentially very large set of auxiliary degrees of freedom from the problem.

\section{Mapping QCMDO to QUBO}
The general form of a complex QCMDO problem is
\begin{align} \label{QCMDO}
 \min_{\substack{[\mathbf{x}]_i \in \mathcal{S}_i \\ \mathbf{F} \mathbf{x} = \mathbf{d}}} \mathbf{x}^\dag \mathbf{A} \mathbf{x} + \mathrm{Re}( \mathbf{x}^\dag \mathbf{b} ) + c .
\end{align}
The objective function is defined by Hermitian $\mathbf{A} \in \mathbb{C}^{n \times n}$ and $\mathbf{b} \in \mathbb{C}^n$.
Linear constraints are defined by $\mathbf{F} \in \mathbb{C}^{m \times n}$ and $\mathbf{d} \in \mathbb{C}^m$.
Each of the variables, $[\mathbf{x}]_i$, is restricted to a set $\mathcal{S}_i$
 that is either $\mathbb{C}$ or a finite subset of $\mathbb{C}$.
Without loss of generality, we partition $\mathbf{x}$ into $n_1$ discrete variables, $[\mathbf{x}_1]_i \in \mathcal{S}_i \subset \mathbb{C}$,
 and $n_2$ continuous variables, $\mathbf{x}_2 \in \mathbb{C}^{n_2}$, with
 compatible block structure induced in $\mathbf{A}$, $\mathbf{b}$, and $\mathbf{F}$:
\begin{equation}
 \mathbf{x} = \left[ \begin{array}{c} \mathbf{x}_1 \\ \mathbf{x}_2 \end{array} \right] , \ \ \ \mathbf{A} =  \left[ \begin{array}{cc} \mathbf{A}_{11} & \mathbf{A}_{12} \\ \mathbf{A}_{21} & \mathbf{A}_{22} \end{array} \right]  , \ \ \ \mathbf{b} = \left[ \begin{array}{c} \mathbf{b}_1 \\ \mathbf{b}_2 \end{array} \right], \notag
\end{equation}
 and $\mathbf{F} = \left[ \begin{array}{cc} \mathbf{F}_1 & \mathbf{F}_2 \end{array} \right]$.
To ensure satisfiability of the linear constraints, we assume that $n_2 \ge m$ and $\mathbf{F}_2$ has linearly independent rows. 
Without this assumption, not all values of $\mathbf{x}_{1}$ are guaranteed to respect the linear constraints. 

We remove the linear constraints in Eq. (\ref{QCMDO}) using
 the singular value decomposition (SVD) of $\mathbf{F}_2$,
\begin{align} \label{SVD}
 \mathbf{F}_2 &= \mathbf{U} \left[ \begin{array}{cc} \mathbf{D} & 0 \end{array} \right] \left[ \begin{array}{cc} \mathbf{V} & \overline{\mathbf{V}} \end{array} \right]^\dag = \mathbf{U} \mathbf{D} \mathbf{V}^\dag ,
\end{align}
 where $\mathbf{U}$ and $\left[ \begin{array}{cc} \mathbf{V} & \overline{\mathbf{V}} \end{array} \right]$
 are unitary, and $\mathbf{D}$ is diagonal positive-definite.
The restricted form of $\mathbf{x}_2$ that satisfies the linear constraints is
\begin{equation}
 \mathbf{x}_2 = \mathbf{x}_{2*} + \overline{\mathbf{V}} \mathbf{x}_{\overline{2}}, \ \ \ \mathbf{x}_{2*} = \mathbf{F}_2^{P} (\mathbf{d} - \mathbf{F}_1 \mathbf{x}_1) ,
\end{equation}
 with a matrix pseudoinverse, $\mathbf{F}_2^{P} = \mathbf{V} \mathbf{D}^{-1} \mathbf{U}^\dag$, denoted by `$^{P}$'.
The constrained optimization over $\mathbf{x}_2$ is reduced to an unconstrained optimization
 over $\mathbf{x}_{\overline{2}} \in \mathbb{C}^{n_2-m}$,
\begin{align} \label{reduce1}
 & \min_{\substack{[\mathbf{x}_1]_i \in \mathcal{S}_i \\ \mathbf{x}_{\overline{2}} \in \mathbb{C}^{n_{2}-m}}} \overline{\mathbf{x}}^\dag \overline{\mathbf{A}} \overline{\mathbf{x}} + \mathrm{Re}(\overline{\mathbf{x}}^\dag \overline{\mathbf{b}}) + \overline{c}, \ \ \ \overline{\mathbf{x}} = \left[ \begin{array}{c} \mathbf{x}_1 \\ \mathbf{x}_{\overline{2}} \end{array} \right] , \\
 & \mathbf{W} = \left[ \begin{array}{cc} \mathbf{I} & 0 \\ -\mathbf{F}_2^{P} \mathbf{F}_1 & \overline{\mathbf{V}} \end{array} \right] , \ \ \
 \overline{\mathbf{b}} = \mathbf{W}^\dag \mathbf{b} + 2 \mathbf{W}^\dag \mathbf{A} \left[ \begin{array}{c} 0 \\ \mathbf{F}_2^{P} \mathbf{d} \end{array} \right] , \notag \\
 & \overline{\mathbf{A}} = \mathbf{W}^\dag \mathbf{A} \mathbf{W} , \ \ \ \overline{c} = c + (\mathbf{F}_2^{P} \mathbf{d})^\dag \mathbf{A}_{22} \mathbf{F}_2^{P} \mathbf{d} + \mathrm{Re}(\mathbf{b}_2^\dag \mathbf{F}_2^{P} \mathbf{d} ) \notag ,
\end{align}
 with induced `$_1$' and  `$_{\overline{2}}$' block structure in $\overline{\mathbf{A}}$ and $\overline{\mathbf{b}}$.

For the minimization over $\mathbf{x}_{\overline{2}}$ to be bounded, $\overline{\mathbf{A}}_{\overline{2}\overline{2}}$ must be positive semidefinite.
If $\overline{\mathbf{A}}_{\overline{2}\overline{2}}$ has a nullspace, it must be orthogonal to $(\overline{\mathbf{b}}_{\overline{2}} + 2 \overline{\mathbf{A}}_{\overline{2}1} \mathbf{x}_1)$ for all $\mathbf{x}_1$.
The nullspace then has no effect on the optimization and we choose the minimizer $\mathbf{x}_{\overline{2}} = - \overline{\mathbf{A}}_{\overline{2}\overline{2}}^{P}( \tfrac{1}{2}\overline{\mathbf{b}}_{\overline{2}} + \overline{\mathbf{A}}_{\overline{2}1} \mathbf{x}_1)$.

A natural midpoint of the mapping is the remaining quadratic unconstrained discrete optimization over $\mathbf{x}_1$,
\begin{align} \label{QUDO}
 &\min_{[\mathbf{x}_1]_i \in \mathcal{S}_i} \mathbf{x}_1^\dag \mathbf{H} \mathbf{x}_1 + \mathrm{Re}(\mathbf{x}_1^\dag \mathbf{g}) + f , \ \ \
  f = \overline{c} - \tfrac{1}{4} \overline{\mathbf{b}}_{\overline{2}}^\dag \overline{\mathbf{A}}_{\overline{2}\overline{2}}^{P} \overline{\mathbf{b}}_{\overline{2}}, \notag \\
 & \mathbf{g} = \overline{\mathbf{b}}_1 - \overline{\mathbf{A}}_{1\overline{2}} \overline{\mathbf{A}}_{\overline{2}\overline{2}}^{P} \overline{\mathbf{b}}_{\overline{2}}, \ \ \ 
 \mathbf{H} = \overline{\mathbf{A}}_{11} - \overline{\mathbf{A}}_{1\overline{2}} \overline{\mathbf{A}}_{\overline{2}\overline{2}}^{P} \overline{\mathbf{A}}_{\overline{2}1}.
\end{align}
Assuming the standard cost of dense linear algebra, mapping Eq. (\ref{QCMDO}) to Eq. (\ref{QUDO})
 requires $\mathcal{O}(n_2 n^2)$ operations.

The discrete-to-binary mapping is less straightforward.
We only consider linear maps of the form $\mathbf{x}_1 = \mathbf{x}_{1*} + \mathbf{T} \mathbf{s}$
 for $\mathbf{s} \in \mathbb{B}^p$ that decompose into $[\mathbf{x}_1]_i = [\mathbf{x}_{1*}]_i + \mathbf{t}_i^\dag \mathbf{s}_i$
 for $\mathbf{s}_i \in \mathbb{B}^{p_i}$, where each discrete variable has its own binary subvector.
The best-case scenario of this form is when each choice of subvector $\mathbf{s}_i$ produces a valid element of $\mathcal{S}_i$.
An example of this is when $\mathcal{S}_i$ is $2^{p_i}$ evenly-spaced numbers with spacing $a$ and $[\mathbf{t}_i]_j = 2^{j-1} a$.
The worst-case scenario is when there is one binary variable for each distinct element of $\mathcal{S}_i$
 with $\{ [\mathbf{t}_i]_j : 1 \le j \le p_i \} = \mathcal{S}_i$ and $[\mathbf{x}_{1*}]_i=0$, and
 a penalty is needed to enforce $\| \mathbf{s}_i \|_1 = 1$.
These two cases set bounds, $\lceil \log_2 |\mathcal{S}_i| \rceil \le p_i \le |\mathcal{S}_i|$.
Other efficient mappings may be possible.

In terms of Eq. (\ref{QUDO}), the final QUBO form of Eq. (\ref{QCMDO}) is
\begin{align}
& \min_{\mathbf{s} \in \mathbb{B}^p} \mathbf{s}^{T} \mathbf{M} \mathbf{s} + k, \ \ \ k = f + \mathbf{x}_{1*}^\dag \mathbf{H} \mathbf{x}_{1*} + \mathrm{Re}(\mathbf{x}_{1*}^\dag \mathbf{g}), \label{eq:FinalQUBO}\\
 & \ \ \ [\mathbf{M}]_{ij} = \mathrm{Re}( [\mathbf{T}^\dag \mathbf{H} \mathbf{T}]_{ij} + \delta_{ij} [\mathbf{T}^\dag (\mathbf{g}+ 2 \mathbf{H} \mathbf{x}_{1*})]_i ) . \notag
\end{align}
In the worst-case scenario mentioned above, for each block constrained to $\| \mathbf{s}_i \|_1 = 1$ one may add a block diagonal penalty to $\mathbf{M}$ with $\lambda$ on the off-diagonals and -$\lambda$ on the diagonals, and add $\lambda$ to $f$ where $\lambda = 2 p \|\mathbf{M}\|_2$ for the unpenalized $\mathbf{M}$. 
Note that the optimal value of the objective functional is the same for Eqs. (\ref{QCMDO}) and (\ref{eq:FinalQUBO}). 
Because QCMDO contains QUBO, it is at least as hard as QUBO, which is NP-hard.

The $\mathbf{s} \in \mathbb{B}^p$ solution to the QUBO problem in Eq. (\ref{eq:FinalQUBO}) can be encoded
 as a computational basis state, $|\mathbf{s}\rangle$, in a $p$-qubit Hilbert space.
$|\mathbf{s}\rangle$ is the final ground state of an AQO
 Hamiltonian with linear $\sigma_x$ couplings and both linear and quadratic $\sigma_z$ couplings,
\begin{align} \label{adiabatic}
 H(t) & = [w(t) - 1] \sum_i \sigma_x^i \\ & \ \ \  + \frac{w(t)}{\Lambda} \left[  \sum_i [\mathbf{M} \mathbf{1} ]_i \sigma_z^i + \sum_{i<j} [\mathbf{M}]_{ij} \sigma_z^i \otimes \sigma_z^j \right] \notag , \\
 \Lambda &= \max \left\{ \max_i | [\mathbf{M} \mathbf{1} ]_i | , \max_{i<j} | [\mathbf{M}]_{ij} | \right\}. \notag
\end{align}
We initialize the AQO to $|+\rangle^{\otimes p}$, for $|+\rangle = \tfrac{1}{\sqrt{2}}( |0\rangle + |1\rangle)$,
 and monotonically increase $w(t)$ from 0 to 1 as the time, $t$, is varied from $t_{\mathrm{initial}}$ to $t_{\mathrm{final}}$.
The maximum coupling strength is normalized to a unit magnitude corresponding to the largest realizable coupling within the AQO.
The ground state energy gap is initially 2, and the runtime, $t_{\mathrm{final}} - t_{\mathrm{initial}}$, depends on the minimum gap,
 the desired accuracy, and the form of $w(t)$ \cite{Lidar2009}.

\subsection{PDE-constrained combinatorial optimization}
An important class of QCMDO problems are those with constraints derived from a discretized linear PDE or system of linear PDEs. 
Accurate discretization of the PDE often requires a large number of auxiliary variables, and it is desirable to perform a dimensional reduction that eliminates these variables from the problem. 
We term the following subclass of QCMDO problems PDE-constrained combinatorial optimization, in analogy with the field from which such problems frequently arise. 
However, a problem of this type may also originate from a non-PDE linear constraint. 

We consider the case where a small observation vector, $\mathbf{x}_{2a} \in \mathbb{C}^{n_{2a}}$, is to be optimized to match a design vector, $\mathbf{y}$, relative to a positive definite metric matrix, $\mathbf{G}$,
\begin{equation}\label{metric_fit}
 \min_{\mathbf{x}_{2a}} (\mathbf{x}_{2a} - \mathbf{y})^\dag \mathbf{G} (\mathbf{x}_{2a} - \mathbf{y}).
\end{equation}
The observation vector is related to a field vector, $\mathbf{x}_{2b} \in \mathbb{C}^{n_{2b}}$,
 through a measurement matrix, $\mathbf{x}_{2a} = \mathbf{K}^\dag \mathbf{x}_{2b}$.
The field vector satisfies a PDE constraint, $\mathbf{E} \mathbf{x}_{2b} = \mathbf{f} + \mathbf{J} \mathbf{x}_1$,
 where $\mathbf{E}$ is an invertible discretized PDE operator,
 $\mathbf{f}$ are uncontrollable boundary values, and $\mathbf{J}$ linearly relates the discrete controls to the controllable boundary values.

This problem written in the form of Eq. (\ref{QCMDO}) is
\begin{align}
 \mathbf{x} &= \left[ \begin{array}{l} \mathbf{x}_1 \\ \mathbf{x}_{2a} \\ \mathbf{x}_{2b} \end{array} \right] , \ \ \
 \mathbf{A} = \left[ \begin{array}{ccc} 0 & 0 & 0 \\ 0 & \mathbf{G} & 0 \\ 0 & 0 & 0 \end{array} \right] , \ \ \
 \mathbf{b} = \left[ \begin{array}{c} 0 \\ - 2 \mathbf{G} \mathbf{y} \\ 0 \end{array} \right] , \notag \\
 \mathbf{F} &= \left[ \begin{array}{ccc} 0 & \mathbf{I} & -\mathbf{K}^\dag \\ -\mathbf{J} & 0 & \mathbf{E} \end{array} \right], \ \ \
 \mathbf{d} = \left[ \begin{array}{c} 0 \\ \mathbf{f} \end{array} \right] , \ \ \
 c = \mathbf{y}^\dag \mathbf{G} \mathbf{y}.
\end{align}
Since $\mathbf{E}$ is invertible, $\mathbf{F}_{2}$ has no nullspace and terms containing $\overline{\mathbf{V}}$ do not appear. 
In this notation, the continuous variable block, `$_2$', is split into `$_{2a}$' and `$_{2b}$' with $n_{2b} \gg n_{2a}$.
For large $n_{2b}$, greater-than-linear costs in $n_{2b}$ are often infeasible.
However, the unconstrained form of Eq. (\ref{QUDO}) is simple for this class,
\begin{align} \label{PDECCO}
 f &= ( \mathbf{y} - \mathbf{K}^\dag \mathbf{E}^{-1} \mathbf{f} )^\dag \mathbf{G} ( \mathbf{y} - \mathbf{K}^\dag \mathbf{E}^{-1} \mathbf{f} ) \\
   &+ (\mathbf{E}^{-1} \mathbf{f})^{\dag} \mathbf{G} (\mathbf{E}^{-1} \mathbf{f}), \notag \\
 \mathbf{g} &= 2 (\mathbf{K}^\dag \mathbf{E}^{-1} \mathbf{J})^\dag \mathbf{G} ( \mathbf{K}^\dag \mathbf{E}^{-1} \mathbf{f} - \mathbf{y} ) , \notag \\
 \mathbf{H} &= (\mathbf{K}^\dag \mathbf{E}^{-1} \mathbf{J})^\dag \mathbf{G} (\mathbf{K}^\dag \mathbf{E}^{-1} \mathbf{J}) . \notag
\end{align}
Assuming that $\mathbf{E}$ is sparse or structured, $\mathbf{E}^{-1} \left[ \begin{array}{cc} \mathbf{f} & \mathbf{J} \end{array} \right]$
 can be calculated efficiently using iterative linear solvers in $\mathcal{O}(n_1 n)$ operations per iteration
 and with few iterations if a good preconditioner is known.
The remaining algebra needs $\mathcal{O}(n_1 n_{2a} n + n_{2a}^3)$ operations.

If $n_1 \le n_{2a}$ and we are free to choose $\mathbf{G}$, then
\begin{equation}
 \mathbf{G} = (\mathbf{K}^\dag \mathbf{E}^{-1} \mathbf{J})^{P \dag} \overline{\mathbf{D}} (\mathbf{K}^\dag \mathbf{E}^{-1} \mathbf{J})^{P}
\end{equation}
for any positive definite diagonal $\overline{\mathbf{D}}$ produces a diagonal $\mathbf{H}$ that reduces the problem to a trivial independent optimization over each discrete variable.

\subsubsection{PDE-constrained QCMDO is NP-hard}
Though the mapping from QCMDO to QUBO is more efficient in the PDE-constrained case,
 this does not alter the computational complexity of the resulting QUBO problem.
To prove this assertion, we show that the NP-hard \textsc{Max-Cut} problem efficiently reduces to a problem of the form in Eq. (\ref{metric_fit}). 

First, we show that \textsc{Max-Cut} can be expressed as an instance of QUBO. 
Given a graph $\mathcal{G}$ with $n$ vertices, the graph Laplacian $\mathbf{L}$ is an $n \times n$ positive-semidefinite matrix equal to the degree matrix of $\mathcal{G}$ minus its adjacency matrix \cite{Merris1994}. 
It is simple to show that the problem $\max_{\mathbf{x} \in \lbrace -1, 1 \rbrace^{n}} \mathbf{x}^{T} \mathbf{L} \mathbf{x}$ is equivalent to the \textsc{Max-Cut} problem, which is NP-hard \cite{Garey1976}. 
We can transform this problem from a maximization into a minimization trivially by taking $\mathbf{L} \to -\mathbf{L}$. 
Let $d$ denote the maximum vertex degree of the graph $\mathcal{G}$. 
Then, adding the diagonal matrix $2 d \mathbf{I}$ to $-\mathbf{L}$ makes $\mathbf{Q} = 2 d \mathbf{I} - \mathbf{L}$ positive semidefinite, since the eigenvalues of $\mathbf{L}$ are bounded above by $2 d$ \cite{Anderson1985}. 
This modification introduces a constant offset, $\mathbf{x}^T (2 d\mathbf{I}) \mathbf{x} = 2 d n$, and does not alter the minimizing $\mathbf{x}$.
Letting `$\equiv$' denote equivalent problems, 
\bea
\textrm{\textsc{Max-Cut}} & \equiv & \min_{\mathbf{x} \in \lbrace -1, 1 \rbrace^{n}} \mathbf{x}^{T} (-\mathbf{L}) \mathbf{x} \nonumber \\
& \equiv &  \min_{\mathbf{x} \in \lbrace -1, 1 \rbrace^{n}} \mathbf{x}^{T} \mathbf{Q} \mathbf{x} \nonumber \\
& \equiv &  \min_{\mathbf{s} \in \mathbb{B}^{n}} (2 \mathbf{s} - \mathbf{1})^{T} \mathbf{Q} (2 \mathbf{s} - \mathbf{1}) \nonumber \\
& \equiv &  \min_{\mathbf{s} \in \mathbb{B}^{n}} \mathbf{s}^{T} \mathbf{Q} \mathbf{s} + \mathbf{s}^{T} \mathbf{v},
\eea
where $\mathbf{1}$ is the vector of all ones and $\mathbf{v} = -\mathbf{Q} \mathbf{1}$. 
To show that for any positive semidefinite $\mathbf{Q}$ and arbitrary $\mathbf{v}$ this problem efficiently reduces to some PDE-constrained QCMDO problem, we simply consider a case where the interior domain governed by the PDE is reduced in size and eventually eliminated, leaving only Dirichlet boundary conditions applied to a vector of boundary points. 
In this case the matrix representation of the PDE is trivial, with $\mathbf{E} = \mathbf{K} = \mathbf{I}$. 
We take the metric to be $\mathbf{G} = \mathbf{I}$. 
If we let $\mathbf{J} = \mathbf{Q}^{1/2}$, then we achieve the unconstrained form of the PDE-constrained QCMDO problem in Eq. (\ref{PDECCO}),
\be
\min_{\mathbf{s} \in \mathbb{B}^{n}} \mathbf{s}^{T} \mathbf{Q} \mathbf{s} + \mathbf{s}^{T} \mathbf{v} \equiv \min_{\mathbf{s} \in \mathbb{B}^{n}} \mathbf{s}^{T} (\mathbf{J}^{\dagger} \mathbf{J}) \mathbf{s} + \mathrm{Re}( \mathbf{s}^{T} \mathbf{g} ),
\ee
for $\mathbf{f}$, $\mathbf{y}$ chosen such that $\mathbf{J} \mathbf{1} = 2 ( \mathbf{y} - \mathbf{f} )$. 
We have shown that the \textsc{Max-Cut} problem for any given graph $\mathcal{G}$ can be efficiently mapped to a corresponding PDE-constrained QCMDO problem. 
Consequently, PDE-constrained QCMDO is NP-hard.

\section{Example \label{sec:Example}}
To illustrate the QCMDO-to-QUBO mapping with a PDE constraint, we consider a potential, $V(\vec{x})$, on a square domain, $\vec{x} \in [0, N]^2$, governed by the Poisson equation, $\nabla^2 V = \rho$, with $V=0$ on the boundary.
The charge density is constrained to a set of Gaussians positioned on a sublattice of a regular lattice,
\begin{align}
 \rho(\vec{x}) &= \sum_{i=1}^{2 N^2 - 2 N +1} [\mathbf{s}]_i \frac{25}{\pi} \exp( - 25 | \vec{x}- \vec{y}_i|^2 ), \\
 \vec{y}_{i+Nj} &= (i-0.5,j-0.5) , \ \ \ i,j \in \{1, ... , N\}, \notag \\
 \vec{y}_{N^2+i+Nj} &= (i,j), \ \ \ i,j \in \{1, ... , N-1\} . \notag
\end{align}
Our goal is to determine $\mathbf{s}$
 by measuring eigenmodes of the potential, $\phi_{m+Nn}(\vec{x}) = \sin(m \pi x_1 / N) \sin(n \pi x_2 / N)$,
 with eigenvalues $\lambda_{m+Nn} = -\pi^2 (m^2 + n^2)/N^2$ for $m,n \in \{1, ... , N\}$.
For $[\mathbf{s}]_i \in \mathbb{R}$, this problem is underdetermined.
The discrete nature of $\mathbf{s}$ is necessary for reconstruction of $\rho(\vec{x})$ from an incomplete set of measurements.

\begin{figure}[h]
\begin{center}
\includegraphics[width=0.45\textwidth]{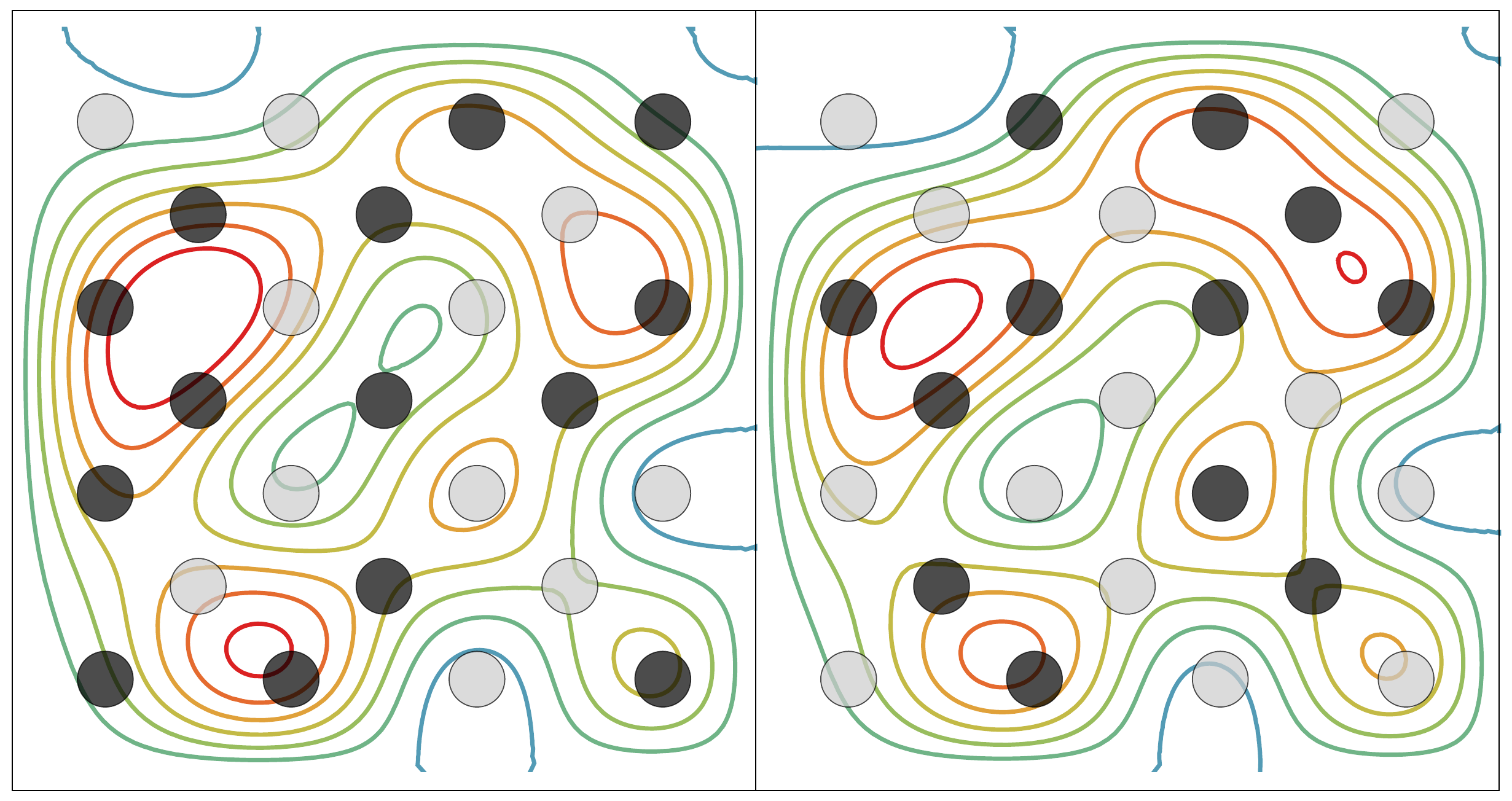}
\end{center}
\vspace{-10pt}
\caption{(Color online) Charges (black and grey dots denote the presence and absence of a charge) and equipotential contours of the measured potential
 for the optimal (left) and best sub-optimal (right) solutions of a charge reconstruction example.}
\label{fig:PoissonEqnExample}
\end{figure}

To put this example into the form of Eq. (\ref{PDECCO}), we discretize the potential on a square grid with a grid spacing of $0.1$
 and use a spectral representation of $\mathbf{E}$,
\begin{align} \label{electrostatics}
 [\mathbf{J}]_{ij} &= 0.1 \frac{25}{\pi} \exp( - 25 | \vec{x}_i - \vec{y}_j|^2 ) , \\ 
 [\mathbf{K}^\dag \mathbf{E}^{-1}]_{ij} &= 0.1 \lambda_i^{-1} \phi_i(\vec{x}_j), \ \ \  \mathbf{f} = 0, \ \ \ \mathbf{G}=\mathbf{I}, \notag \\ 
 \vec{x}_{i+Nj} &= (0.1 i, 0.1 j), \ \ \ i,j \in \{1,10 N - 1\}. \notag
\end{align}
We examine a case for $n_{2a} = 16$, $n_{2b} =1521$, and $p = 25$
 with randomly assigned charges.
In Fig. \ref{fig:PoissonEqnExample} we plot the measured potentials and charge configurations
 for an optimal (ground state) and best sub-optimal solution (first excited state).
This example is chosen for its large Hamming distance of 16 between the $\mathbf{s}$ vectors of the ground and excited states.
The measured potential is constructed by summing the measured eigenmodes with their coefficients.
The similarity between potentials of distinct charge distributions is the result of an effective low-pass filtering
 and highlights the combinatorial difficulty of this optimization problem.

\begin{figure}[h]
\begin{center}
\includegraphics[width=0.45\textwidth]{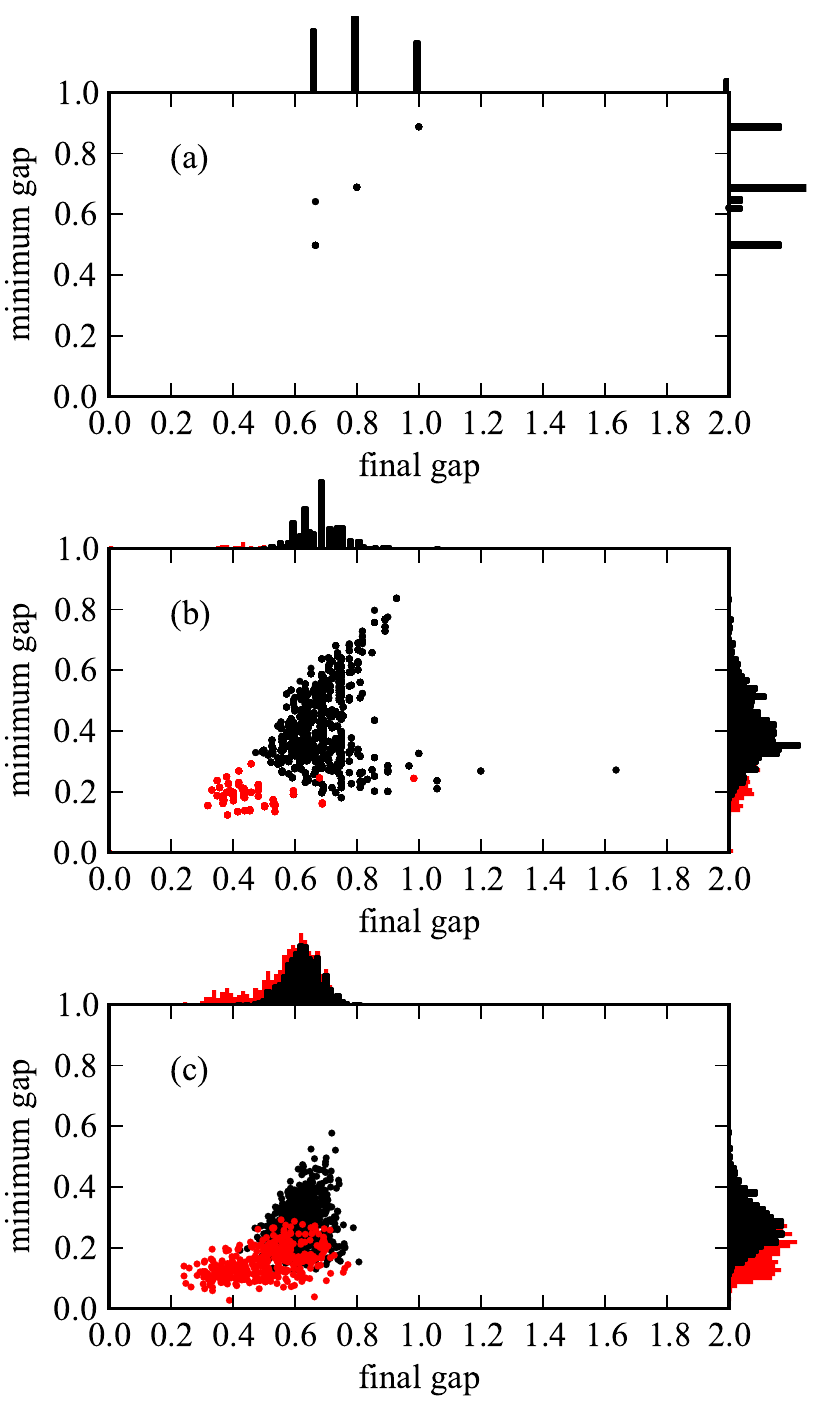}
\end{center}
\vspace{-10pt}
\caption{(Color online) Distribution and correlation of the minimum and final gaps of the electrostatics example in Eq. (\ref{electrostatics}) as implemented in the AQO Hamiltonian in Eq. (\ref{adiabatic}) for (a) all instances of $p = 5$, (b) all instances of $p = 13$, and (c) 1000 random samples of $p = 25$. 
Red data points indicate cases when the Hamming distance between the ground and first excited state charge distributions is greater than 1.}
\label{fig:Statistics}
\end{figure}

\section{AQO implementation issues}
The difficulty of an optimization problem in an AQO implementation grows with the inverse of the minimum energy gap
 of Eq. (\ref{adiabatic}) along the adiabatic path.
In addition, the final energy gap determines how sensitive the global optimum is to errors in the couplings.
A disparity between these gaps indicates a numerically well-defined global optimum that is difficult to prepare through the prescribed adiabatic path.
We simulate the AQO implementation of Eq. (\ref{electrostatics}) in the Hamiltonian of Eq. (\ref{adiabatic})
 with exact diagonalization on a classical computer with results summarized in Fig. \ref{fig:Statistics}.
Samples are organized into two populations according to whether or not the Hamming distance between
 the ground and first excited state charge distributions is $1$.
Observed non-unity Hamming distances range from $6$ to $16$, with larger Hamming distances correlating well with smaller minimum and final gaps.
For the example shown in Fig. (\ref{fig:PoissonEqnExample}), the final gap is $0.483$ and the minimum gap is $0.096$.

The matrix $\mathbf{M}$ parameterizing the quadratic term of the QUBO objective function is dense, in general. 
In order to implement this QUBO problem on an AQO, the complete coupling graph describing the problem must be embedded into a hardware-realizable coupling graph. 
This problem is known as minor embedding \cite{Choi2008, Choi2011}. 
Hardware limitations typically include constraints on the spatial locality of the couplings and on the degree of the coupling graph. 
The embedding step leads to an overhead in both the number of qubits required \cite{Choi2011} and the strength (or equivalently, precision) of the qubit-qubit couplings \cite{Choi2008}. 
For the hardware graph implemented on the D-Wave device \cite{Harris2010}, for example, embedding a complete graph incurs a quadratic overhead in the number of qubits \cite{Choi2011, Klymko2014}. 
We note that the largest problem size of 25 qubits that we have studied with exact diagonalization in Sec. \ref{sec:Example} may be embedded into the Chimera graph of the 512 qubit D-Wave 2 device, using the complete graph embedding algorithm of Klymko, et al. \cite{Klymko2014}. 

A necessary condition for the Hamiltonian embedding to be successful is that the ground state of the Hamiltonian as implemented be equivalent to the solution of the original QUBO problem. 
However, finite precision in the coupling parameters may lead to errors of the form of implementing a perturbed Hamiltonian with a different ground state. 
In addition, during the quantum annealing process excitations due to non-adiabaticity and coupling to the environment will lead to suppressed occupation of the ground state. 
As a result, the annealing step may need to be repeated many times in order to obtain a sufficiently large probability of measuring the optimal solution \cite{Boixo2014}.

\section{Conclusion}
In this work, we have extended the class of problems that may be implemented on an adiabatic quantum optimizer (AQO) to include quadratic constrained mixed discrete optimization (QCMDO). 
QCMDO corresponds to those optimization problems for which the objective functional is quadratic, the constraints are linear, and the optimization parameters are a mix of continuous and discrete controls. 
We construct an efficient dimension-reducing mapping from any given QCMDO problem to a quadratic unconstrained binary optimization (QUBO) problem, which may then be implemented on an existing AQO. 
Included in the class of QCMDO problems are those for which the linear constraint is given by a linear partial differential equation (PDE) or system of linear PDEs. 
This mapping is suitable for use by either an AQO or a standard classical solver. 

\section*{Acknowledgements}
We thank Ojas Parekh and Denis Ridzal for informative discussions. 
This work was supported by the Laboratory Directed Research and Development program at Sandia National Laboratories. 
Sandia National Laboratories is a multi-program laboratory managed and operated by Sandia Corporation, a wholly owned subsidiary of Lockheed Martin Corporation, for the U.S. Department of Energy National Nuclear Security Administration under contract DE-AC04-94AL85000.
\appendix
\bibliography{mybibliography}
\end{document}